\documentclass[aps,prl,twocolumn,superscriptaddress]{revtex4}
\usepackage{graphicx,amssymb,amsmath}

\begin{document}

\title{Fermi Arcs in the Superconducting Clustered State for Underdoped Cuprates}
\author{G. Alvarez}
\affiliation{Computer Science \& Mathematics 
Division and Center for Nanophase Materials Sciences, Oak Ridge National Laboratory, \mbox{Oak Ridge, TN 37831}}
\author{E. Dagotto}
\affiliation{Department of Physics and Astronomy, University of Tennessee, Knoxville, Tennessee 37996, USA, and \\
Materials Science and Technology Division, Oak Ridge National Laboratory, Oak Ridge, Tennessee 32831, USA}

\begin{abstract}
The one-particle spectral function of a state formed by superconducting (SC) clusters 
is studied via Monte Carlo techniques. The clusters have similar SC amplitudes but randomly
distributed phases. This state is stabilized by the 
competition with antiferromagnetism,
after quenched disorder is introduced. Fermi arcs between the critical
temperature $T_{\rm c}$ and the cluster formation 
temperature scale $T^*$ are observed, similarly
as in the pseudogap state of the cuprates. The
arcs originate at metallic regions in between the
neighboring clusters that present large SC phase
differences. 
\end{abstract}

\maketitle

{\it Introduction.}
Research in hole-doped high temperature superconductors is currently mainly focusing on the
pseudogap (PG) phase that exists above $T_{\rm c}$ in the underdoped regime. This exotic
phase acts as the normal state to superconductivity in a broad range of hole densities, and it may
contain the solution to the puzzling properties of these compounds. Two groups
of experiments have recently provided important microscopic information about the PG phase: {\it (i)}
Using angle-resolved photoemission (ARPES) techniques~\cite{damascelli}, 
the PG Fermi surface was found to be composed of
disconnected segments, widely known as ``Fermi arcs''~\cite{norman}, that are  
centered at the nodal (N) points ${\bf k}$=$(\pm \pi/2,\pm \pi/2)$, in the
usual two-dimensional (2D) square-lattice notation.
These arcs are caused by the low-energy PG, which presents a ${\bf k}$-dependence resembling 
a $d$-wave SC gap close 
to the anti-nodal (AN) points $(0,\pm \pi)$,$(\pm \pi,0)$~\cite{damascelli}, 
but it has
gapless excitations in a finite momentum range near the nodes~\cite{kanigel,kanigel2}.
The AN gaps remain approximately constant in the PG phase, in contrast 
with mean-field theories~\cite{kanigel,kanigel2}. {\it (ii)} Using scanning tunneling microscopy (STM) techniques,
remarkable results were recently reported~\cite{yazdani}. 
At temperatures well above $T_{\rm c}$,
the B2212 local density-of-state (LDOS) still closely resembles 
that of a $d$-wave superconductor. E.g., at optimal doping
and $T$=120K, 
a $d$-wave gap is observed 
in clusters that appear to be randomly distributed~\cite{nernst}.

Theoretically, the PG state is believed to be either {\it (i)} a precursor of the SC state, with
phase fluctuations destroying superconductivity in an homogeneous state made out of small Cooper pairs   
formed by a strong attraction~\cite{randeria}, or {\it (ii)} caused by other competing orders. However,
a third possibility was recently proposed~\cite{alvarez}. Via
Landau-Ginzburg (LG) calculations,
a state with nano-scale SC and antiferromagnetic (AF) clusters is stabilized when 
quenched disorder (caused, e.g., by chemical doping) influences 
on an otherwise homogeneous state with
local coexistence of the SC and AF order parameters.
The SC amplitude is robust and approximately
the same in all the SC clusters. However, the phase of the SC order parameter, while 
uniform within each SC cluster, randomly changes from cluster to cluster causing the overall
state to become non-SC.
This scenario is conceptually different from {\it (i)} and {\it (ii)}, but contains elements of both:
the AF competing order is needed for the SC clusters to form and, once formed, phase
fluctuations between clusters, occurring even in a weak-coupling BCS regime, destroy the SC condensate.
This ``Superconducting Clustered State'' (SCCS) can account, at least qualitatively, for
the STM experiments~\cite{yazdani}, since both are based on a similar 
inhomogeneous distribution of
SC gaps above $T_{\rm c}$. In this manuscript, it is shown that the SCCS also produces a Fermi surface (FS)
with Fermi arcs. 
\begin{figure}
\centerline{
\includegraphics[width=7.2cm,clip]{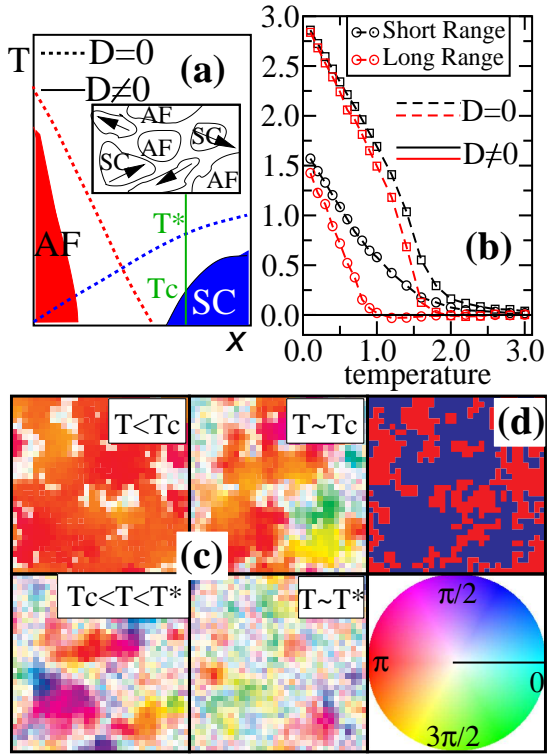}}
\caption{(Color online) (a) Schematic $T$ vs. doping ($x$) phase diagram~\cite{alvarez} 
for the competition AF vs. SC: $D$=0 ($D$$\neq$0) is the clean (dirty) limit.
The vertical (green) line is the $T$-range emphasized here: at $T_{\rm c}$ 
long-range order develops, and at $T^*$ SC clusters (short-range order) are formed.
In practice, the AF to SC transition is reached in Eq.(1) by varying $\rho_{\rm SC}$ and $\rho_{\rm AF}$,
keeping $\rho_{\rm AF}$=1+$\rho_{\rm SC}$ for simplicity~\cite{alvarez}. 
The disorder in the couplings is $correlated$ with power $\alpha$=0.8~\cite{correlated}.
(b) SC correlation (SS) vs. $T$ at short (vectorial distance (3,0)) 
and long (vectorial distance (16,16))  
distance, in the clean and dirty limits. Results are from MC simulations on a 32$\times$32 cluster 
with periodic boundary conditions, using
2,000 (3,000) thermalization (measurement) steps, starting with a random initial configuration. $\rho_{\rm SC}$
was taken from a bimodal distribution with values -1.1 and -0.1 in the dirty limit, and uniformly -1.1 in the clean limit.
SS is defined as 
$SS({\bf i})$=$\frac{1}{N_{sites}}\sum_{{\bf j}} |\Delta_{\bf j}| |\Delta_{\bf j+i}| \cos(\phi_{\bf j}-\phi_{\bf j+i})$.
Some results gathered on 64$\times$64 lattices show that size effects are not important.
(c) Classical SC order parameter $\Delta$ for a typical MC-equilibrated 
configuration at several $T$'s 
($0.2$, $1.0$$\sim$$T_{\rm c}$, $1.5$ and $2.0$$\sim$$T^*$).
Color intensity represents $|\Delta|$ and  the actual colors represent the SC
angle $\phi$ at each site (see color wheel).
The maximum value of $|\Delta|$ for the temperatures studied was $\sim$2.0.
The bimodal couplings configuration is also shown (d): 
blue indicates regions where SC couplings dominate ($\rho_{\rm SC}$=$-1.1$);
red  where AF couplings dominate ($\rho_{\rm SC}$=$-0.1$). 
}
\label{fig:fig1}
\end{figure}

{\it LG simulations.}
Our calculations start with a 
LG theory for the AF vs. $d$-wave SC competition, studied with
Monte Carlo (MC) techniques. Details were
extensively discussed before~\cite{alvarez} and only a brief summary will be repeated here. 
The LG classical Hamiltonian involves a complex number 
$\Delta_{\bf i}$=$|\Delta_{\bf i}|$$e^{i\phi_{\bf i}}$  
and a real vector ${\bf S}_{\bf i}$,
representing the SC and AF order parameters at site {\bf i} of a 2D square lattice. 
The interactions are the standard: 1. terms with up to 4th powers of the order
parameters, locally favoring SC and AF; 2. nearest-neighbor (NN) couplings that spread the range of the order,
emerging from gradients in the continuum formulation; and 3. an interaction 
between the order parameters, with strength $u_{\rm SC|AF}$. Quenched disorder is also included to represent 
chemically doped cuprates. More specifically,
\begin{eqnarray}\nonumber
H&=&r_{\rm SC} \sum_{\bf i}|\Delta_{\bf i}|^2 +
\frac{u_{\rm SC}}{2}\sum_{\bf i}|\Delta_{\bf i}|^4 \\ \nonumber
&+& \sum_{{\bf i},\alpha}\rho_{\rm AF}({\bf i},\alpha){\bf S}_{\bf i}\cdot {\bf S}_{{{\bf i}}+\alpha}
+u_{\rm SC|AF}\sum_{\bf i} |\Delta_{\bf i}|^2|{\bf S}_{\bf i}|^2\\  \nonumber
&+&\sum_{{\bf i},\alpha} \rho_{\rm SC}({\bf i},\alpha)|\Delta_{\bf i}||\Delta_{{\bf i}+\alpha}|
\cos(\phi_{\bf i}-\phi_{{\bf i}+\alpha})\\
&+& {\it r_{\rm AF}} \sum_{\bf i} |{\bf S}_{\bf i}|^2 + \frac{u_{\rm AF}}{2} \sum_{\bf i} |{\bf S}_{\bf i}|^4.  
\label{eq:hamgl}
\end{eqnarray} 
Previous investigations~\cite{alvarez} showed that fixing, e.g., 
$r_{\rm SC}= -1$, $r_{\rm AF}$=$-0.85$, $u_{\rm SC}$=$u_{\rm AF}$=$1$, $u_{\rm SC|AF}$=$0.7$ 
and varying $\rho_{\rm SC}$ and $\rho_{\rm AF}$ along the line $\rho_{\rm AF}$=1+$\rho_{\rm SC}$,
produces a clean-limit phase diagram with a SC+AF region of local coexistence~\cite{alvarez}. 
Fig.~1a qualitatively illustrates this clean limit case ($D$=0)~\cite{alloul}. 
However, quenched disorder $D$$\neq$0, for example 
introduced in Eq.(1) via a random distribution of $\rho_{\rm SC}$ and
$\rho_{\rm AF}$ values, reduces both critical temperature, 
opening a gap between the competing phases. In this glassy region, 
nano-scale SC and AF clusters coexist, as sketched
in the inset~\cite{alvarez}. Two temperature scales emerge naturally: upon cooling,
first the SC amplitude develops at a crossover scale $T^*$, but the SC phases 
between clusters remain random. 
Reducing $T$ further, coherence among the cluster phases is reached at $T_{\rm c}$. The presence of
two temperature scales is indeed observed numerically, as shown in Fig.1b: at $D$=0, the
long and short correlations are very similar, while with disorder $D$$\neq$0, a substantial
difference between $T^*$ and $T_{\rm c}$ (defined by the $T$ where short and long range correlations 
vanish, respectively) is clearly observed. 
MC-equilibrated configurations of Eq.(1) better clarify this issue (Fig.~1c). 
The complex-number SC orders parameters are 
represented by a color and an intensity (see wheel in Fig.~1c). 
At low $T$, the uniform and intense color indicates a robust SC state.
As $T$ increases, phase fluctuations appear near $T_{\rm c}$. In the interesting range
$T_{\rm c}$$<$$T$$<$$T^*$, both SC clusters with random phases 
and non-SC (white) regions coexist: this is the
SCCS emphasized here~\cite{comment2}.
Finally, near $T^*$ or above, few vestiges of SC remain.

{\it Fermions in the SC+AF background.}
After equilibrated configurations of the SCCS state 
are gathered from the LG/MC classical analysis, fermionic properties are obtained by
locally coupling itinerant electrons (simulating carriers) 
to the classical order parameters, as previously discussed~\cite{alvarez}. The
Hamiltonian is:
\begin{eqnarray}\nonumber
&H_{\rm F}&=-t\sum_{<{\bf ij}>,\sigma}(c^\dagger_{{\bf i}\sigma}c_{{\bf
j}\sigma}
+H.c.) + 2 \sum_{\bf i} J_{\bf i} S_{\bf i}^z s_{\bf i}^z \\
& + & 
\frac {1}{2}\sum_{{\bf i},\alpha}V_{\bf i}|\Delta_{{\bf i}\alpha}|^2-
\sum_{{\bf i},\alpha}V_{\bf i}(\Delta_{{\bf i}\alpha}
c_{{\bf i}\uparrow}c_{{\bf i}+\alpha\downarrow}+H.c.), 
\label{eq:hamfermi}
\end{eqnarray} 
\noindent
where $c_{{\bf i}\sigma}$ are fermionic operators, 
$s_{\bf i}^z$=$(n_{{\bf i}\uparrow}-n_{{\bf i}\downarrow})/2$~\cite{comment}, $n_{{\bf
i}\sigma}$ is
the number operator, and
$\Delta_{{\bf i}\alpha}$=$\beta$$|\Delta_{{\bf i}}|$$e^{i\phi_{\bf i}}$
are complex numbers for the SC order parameter
defined now at the links (${\bf i}$,${\bf i}$+$\alpha$)($\alpha$=unit vector
along the $x$ or $y$ directions; $\beta$=1 (-1) for $\alpha$ along $x$ ($y$)). 
For each $\{\Delta_{{\bf i}\alpha}\}$  and $\{ S_{\bf i}^z \}$ 
configuration, the fermionic
sector is exactly diagonalized via library subroutines and any property, static
or dynamic, can be easily obtained.
At $J_{\bf i}$=0, $d$-wave SC is favored since the pairing term involves NN sites, as in any standard
mean-field Bogoliubov-de-Gennes (BdG) approximation~\cite{comment3}. 
The parameters of relevance are $J_{\bf i}$ and $V_{\bf i}$ ($t$
is the energy unit), and they carry a site dependence to easily include
quenched disorder~\cite{comment4}. 

\begin{figure}
\centerline{
\includegraphics[width=7.2cm,clip]{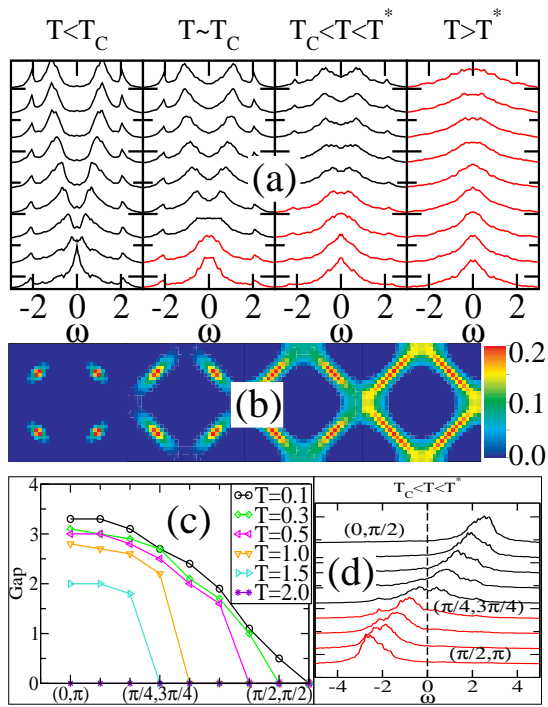}}
\caption{(Color online) (a) $A({\bf k},\omega)$ for equally-spaced ${\bf k}$'s 
along the direction from $(\pi/2,\pi/2)$ (bottom) to $(0,\pi)$ (top). 
The classical LG configurations used are obtained as described in Fig.~1 but 
for the case $\rho_{\rm SC}$=$-0.8$ ($0$) in the SC (AF) regions, 
using $\rho_{\rm AF}$=1+$\rho_{\rm SC}$, on a 32$\times$32 lattice.
The $T$'s are, from left to right, $T$=$0.1$, $0.4$, $1.0$ and $2.0$.
In this case, $T_{\rm c}$$\sim$$0.4$,  and $T^*$$\sim$ 1.2-1.5.
(b) $A({\bf k},\omega$$=0)$ in the $k_x$-$k_y$ plane for the same parameters and temperatures as in (a). Results shown
are those within a window $\Delta \epsilon$=0.1 from the Fermi level.
(c) SC gap (distance between peaks) vs. momentum  for the same parameters as in (a) 
but with $\rho_{\rm SC}$=$-1.1$ or $-0.1$ (bimodal distribution),
at the $T$'s indicated. 
(d) $A({\bf k},\omega$$=0)$ from $(0,\pi/2)$ to $(\pi/2,\pi)$, at $T$=1.0 and parameters as in (a).
}
\label{fig:fig2}
\end{figure}

{\it Fermi arcs.} Fig.~2 contains our most important results. 
(a) shows the one-particle spectral function $A({\bf k},\omega)$
along a straight line from the N to the AN points. At low $T$$<$$T_{\rm c}$, the $d$-wave SC gap is clearly visible
(higher $|\omega|$ peaks are related with a nodeless AF gap, and they do not affect lower energy
features). However, as $T$ is raised first across $T_{\rm c}$, then through 
the intermediate SCCS phase-fluctuating regime proposed for the PG state, and finally to above
$T^*$, clearly the gaps disappear forming segments (arcs) starting at the node, with a length that
grows with increasing $T$.
In (b), the FS's are shown: four nodes at low $T$
become arcs at higher $T$, they eventually merge, and form
a closed FS at the highest $T$. (c) contains an example of gaps
vs. ${\bf k}$ along the N-AN line, showing the arc formation, and the stability of the AN gap even when $T$
is varied over a wide range. Here, as in~\cite{kanigel,kanigel2}, a given ${\bf k}$ is said to have a gap 
if two peaks are found in $A({\bf k},\omega)$.
In Fig.~3a, the length of the Fermi arc is shown vs. $T$, and an approximate linear relation is found.
All these results are in 
good agreement with ARPES~\cite{kanigel,kanigel2}. 

\begin{figure}[h]
\includegraphics[width=7.2cm,clip]{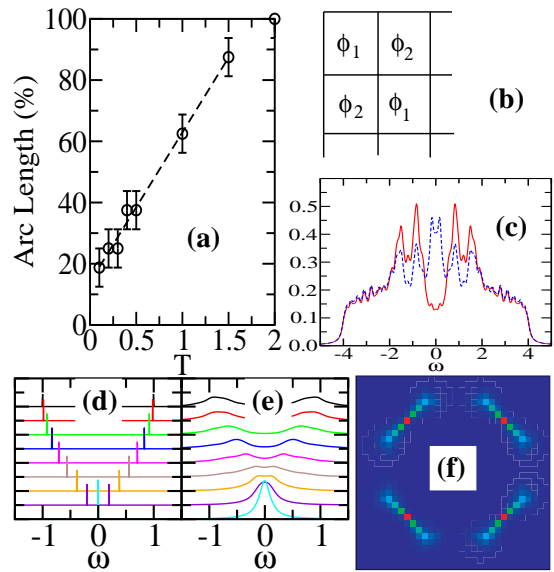}
\caption{(Color online) 
(a) Length of the Fermi arc (as a \% of the maximum length) vs. $T$ for the case described in Fig.~\ref{fig:fig2}a.
The Fermi arc was defined to exist at a certain momentum ${\bf k}$
when its intensity was within 35\% of the maximum intensity. Other definitions lead to a similarly linear relation
but with different $T$$\rightarrow$0 limits. Note that the energy cutoff $\Delta \epsilon$=0.1 and the finite
momentum resolution, due to the lattice's finite size, prevent the size of the nodes from being exactly zero
in the low-$T$ state. If a shift downwards by 20\% (the value at $T$=0) is carried out, an even better agreement
with \cite{kanigel,kanigel2} is obtained. 
(b) Schematic representation of the toy model configuration (see text). $\phi_1$
and $\phi_2$ refer to the SC order parameter phases in 4$\times$4 squares.
(c) LDOS for the example shown in (b) with $\phi_1$=$0$ and $\phi_2$=$\pi$. 
Red (blue) solid (dashed) lines correspond to a site at the center (border) of the 4$\times$4 square. The
parameters used are the same at each site: $|\Delta|$=1, $V$=0.25, and $J$=2.  
(d) $A({\bf k},\omega)$ for ${\bf k}$ along the direction $(\pi/2,\pi/2)$ to $(0,\pi)$ for the 
case shown in (b) with $\phi_1$=$\phi_2=0$ (i.e. perfect $d$-wave superconductor).
(e) Same as (d), but with $\phi_1$ and $\phi_2$ randomly chosen between $0$ and $\pi$.
(f) $A({\bf k},\omega$$=0)$ in the $k_x$-$k_y$ plane for case (e) ($\Delta \epsilon$=0.1).
\label{fig:fig3} }
\end{figure}

An important observation is that the arcs are $not$ merely caused by the broadening of the 
peaks by disorder.
To understand this,
consider now the direction $perpendicular$ to the N-AN line. In Fig.~2d, $A({\bf k},\omega)$
from $(0,\pi/2)$ to $(\pi/2,\pi)$ is shown. This crosses the N-AN line at
$(\pi/4,3\pi/4)$. The figure shows that in the range $T_{\rm c}$$<$$T$$<$$T^*$ 
a $metallic$ dispersion, close to non-interacting electrons,
is observed. This is totally different from the low-$T$ results that show a BdG
quasiparticle dispersion and a gap (not shown). Concomitant with this behavior, by monitoring the LDOS
in our simulations, we have noticed the existence of gapless metallic patches coexisting with the AF and
SC clusters. These metallic regions appear in ``fragile'' zones of the
disordered configuration, such as in long and thin areas of one phase penetrating into the other, where none of
the two orders prevails. 
Thus, the SCCS actually involves
three ingredients: SC, AF, and metallic areas. Our numerical simulations suggest that the
metallic areas and Fermi arcs are related.

{\it Toy models.} To better understand the Fermi arc formation in the SCCS, simplified models
were analyzed. Consider a 2D square lattice regularly divided into smaller 4$\times$4 squares (Fig.~\ref{fig:fig3}b),
all with the same SC amplitude but different phases, and without AF.
To simulate $T$ effects, frozen configurations $\{\phi\}$ of the SC phases 
were studied.
In Fig.~3d, the uniform case $\phi_1$=$\phi_2$=0 is shown: as expected, a clear $d$-wave gap exists along the N-AN line.
This mimics the regime $T$$<$$T_{\rm c}$ in Fig.~\ref{fig:fig2}a. 
To simulate $T_{\rm c}$$<$$T$$<$$T^*$, consider now a {\it random} distribution of phases. The result
(Fig.~\ref{fig:fig3}e) still shows a clear gap in the AN point, but near the nodal point now
a finite set of momenta do {\it not} present a gap anymore, 
thus generating a Fermi arc (see also the FS in Fig.~\ref{fig:fig3}f)). 
Studying $A({\bf k},\omega)$ 
for a variety of $\phi_2$'s, at fixed $\phi_1$=0,
we observed that angles such that $\cos(\phi_2)$$<$0 are those that most contribute to the arcs. 
For, e.g., $\phi_2$=$\pi$ metallic portions were identified 
at the lines separating two domains (Fig.~3c)~\cite{annett}.

\begin{figure}[h]
\includegraphics[width=7.2cm,clip]{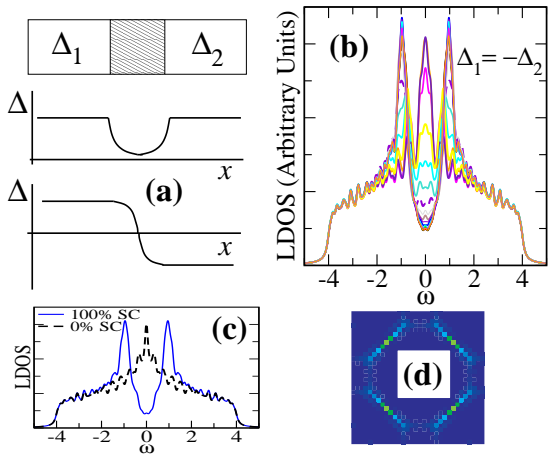}
\caption{(Color online) (a) Schematic representation of a Josephson-junction-like structure. Top: two SC regions
with order parameters $\Delta_1$ and $\Delta_2$, separated by a $\Delta$=0 area of width $w$ (grey). Middle: expected gap 
interpolation due to proximity effect, for $\Delta_1$=$\Delta_2$. 
Bottom: same as middle, but for $\Delta_1$=-$\Delta_2$.
(b)  LDOS on a 32$\times$32 lattice containing 4 equally-spaced 12$\times$12 SC clusters separated by $w$=4.
The order parameters are staggered between $\Delta_1$=1 and $\Delta_2$=-1, with $V$=0.25 and without AF ($J$=0).
Shown are results
from the center of one SC region to a nearest-neighbor.
(c) Density of states of a perfect superconductor (blue solid) and of a perfect metal (black dashed)
on a 32$\times$32 lattice. Note the Van Hove singularity in the 2D metal. 
(d) $A({\bf k},\omega=0)$ in the $k_x$-$k_y$ plane for $\Delta_1$=$-\Delta_2$=1
and $w$=2, with a setup as described in (b) but using 14$\times$14 SC clusters.
\label{fig:fig4}}
\end{figure}

{\it Relevance of large phase differences.} 
The metallic regions seem caused, at least in part, by
large phase differences between neighboring SC clusters. Consider 
Fig.~\ref{fig:fig4}a: here two SC clusters
are shown, with order parameters $\Delta_1$$\neq$0 and $\Delta_2$$\neq$0, separated by an intermediate thin region where $\Delta$=0. 
If $\Delta_1$=$\Delta_2$, this intermediate region will 
develop a gap by proximity effect (Fig.~\ref{fig:fig4}a), and the FS
has still 4 nodes,
as confirmed by an explicit BdG calculation (not shown). 
However, if $\Delta_1$=-$\Delta_2$ a qualitative difference occurs: now 
the interpolation in the intermediate region necessarily  requires the existence of a {\it zero}, where the SC order
parameter must vanish even for a thin $\Delta$=0 layer (Fig.~\ref{fig:fig4}a). A BdG study confirms that the LDOS in the middle
between two ``anti-parallel'' SC clusters is almost identical to the LDOS of 
a metallic state (see Figs.~4b,c).
The FS for $\Delta_1$=-$\Delta_2$ has Fermi arcs (Fig.~4d). 
Thus, large $\phi$ differences induce metallic regions, and those appear to cause the Fermi arcs.

{\it Conclusions.}
A state based on SC clusters with random phases
was here used to calculate the ARPES response of the PG state. Fermi arcs were found
in a $T$ range between $T_{\rm c}$ and the $T^*$ where the clusters start forming upon cooling. 
Toy models illustrate and simplify these results. 
For further progress, our effort must
be supplemented by better
analytic control of the SCCS and by numerical studies using 
microscopic models. The extension of these results to $s$-wave superconducting films should also be pursued~\cite{films}.
In addition,
the STM experiments~\cite{yazdani} not only showed 
a clustered stated in a broad temperature range,
but also unveiled a  regime immediately above  $T_{\rm c}$ where 100\% of the sample had the $d$-wave gap. Adapting
the SCCS to also accommodate this feature will require increasing the attraction $V$ \cite{mayr} and
this direction will be the focus of future investigations.

Work supported by the NSF grant DMR-0706020, the Division of Materials Science and Engineering, U.S. DOE,
under contract with UT-Battelle, LLC,
and by the CNMS, sponsored by the Scientific User Facilities Division, BES-DOE.


\begin{thebibliography}{00}
\bibitem{comment2} 
Figure~1c emphasizes one quenched disorder configuration, 
but calculations have been repeated for several other
configurations and they are all qualitatively similar.
\bibitem{comment} We use Ising spin degrees of freedom (d.o.f.). 
Studies with Heisenberg d.o.f. lead to qualitatively similar 
conclusions, but they are more CPU time consuming.
\bibitem{comment4} Our convention was to choose $V$=-0.25$\rho_{\rm SC}$ (0) at the
sites where $\rho_{\rm SC}$$<$0 ($>$0). Intuitively, in the blue sites of Fig.~1d, the
SC attraction is strong and in the red ones, it is weak. Reciprocally, $J$=2.0$\rho_{\rm AF}$ (0)
if $\rho_{\rm AF}$$>$0 ($<$0).
\end{thebibliography}
\end{document}